\documentclass[preprint,nofootinbib,amssymb,amsfonts]{revtex4-1}

\usepackage{graphicx}
\usepackage{tabularx}
\usepackage{natbib}
\newcommand{\abar}{\bar{a}}

\newcommand{\half}{\mbox{\small{$\frac{1}{2}$}}}
\newcommand{\fourth}{\mbox{\small{$\frac{1}{4}$}}}

\newcommand{\threehalves}{\mbox{\small{$\frac{3}{2}$}}}

\begin{document}
\title{Noneuclidean Tessellations and their relation to Reggie Trajectories}
\author{B. H. \surname{Lavenda}}
\email{info@bernardhlavenda.com}
\homepage{www.bernardhlavenda.com}
\affiliation{Universit$\grave{a}$ degli Studi, Camerino 62032 (MC) Italy}
\begin{abstract}
The coefficients in the confluent hypergeometric equation specify the Regge trajectories and the degeneracy of the angular momentum states. Bound states are associated with real angular momenta while resonances are characterized by complex angular momenta. With a centrifugal potential, the half-plane is tessellated by crescents. The addition of an electrostatic potential converts it into a hydrogen atom, and the crescents into  triangles which may have complex conjugate angles; the angle through which a rotation takes place is accompanied by a stretching. Rather than studying the properties of the wave functions themselves, we study their symmetry groups. A complex angle indicates that the group contains loxodromic elements. Since the domain of such groups is not the disc, hyperbolic plane geometry cannot be used. Rather, the theory of the isometric circle is adapted since it treats all groups symmetrically. The pairing of circles and their inverses is likened to  pairing  particles with their antiparticles which then go one to produce nested circles, or a proliferation of particles. A corollary to Laguerre's theorem, which states that the euclidean angle is represented by a pure imaginary projective invariant, represents the imaginary angle in the form of a real projective invariant.
\end{abstract}
\maketitle

\section{Introduction}

Poincar\'e discovered his conformal models of hyperbolic geometry in an attempt to understand whether there were solutions to the hypergeometric equation of higher periodicities than the then known circular and elliptic functions (of genus $0$ and $1$, respectively). The names, Fuchsian and Kleinian functions, which he coined, belong to a certain class of automorphic functions that live on tiles that tessellate the half-plane or disc, depending on which model is chosen.

The indicial equation is a solution to a differential equation in the neighborhood of a singular point. An equation of second-order can have at most three branch points, which are conveniently taken to be $0$, $1$, and $\infty$. The use of a matrix to describe how an algebraic function is branched had been introduced by Hermite, but it was Riemann who first considered products of such matrices.  Frobenius showed that the hypergeometric equation is completely determined by its exponents at its singular point. If the only effect of analytically continuing two solutions around a singular point is to multipy them by a constant, then the differences in the exponents must all be \textit{integers}, without the solution containing a logarithmic term. In other words, the matrices are rotations about each of the singular points where performing a complete circuit multiplies the solution by a constant factor. We will generalize these \lq rotation\rq\ matrices through real angles to \lq complex\rq\ ones, and in so doing elliptic transformations  will become loxodromic ones.

The monodromy matrices of Riemann are generators of a group, and are either homothetic (magnification) or rotation. These are related to the hyperbolic and elliptic geometries, respectively, both of which preserve the unit circle. According to the Riemann mapping theorem, any arbitrary region bounded by a closed curve can be mapped in a one-to-one fashion onto the interior of a unit disc by an analytic function. Riemann arrived at his theorem as an intuitive conjecture in 1852, and it is hardly comprehensible that it took almost fifty years to  prove it. According to Schwarz, the only biunique analytic mappings of the interior of the unit disc onto itself is a linear, fractional (M\"obius) transform of the hyperbolic or elliptic type; that is, magnification or rotation.

The relevant equation for quantum theory is the confluent hypergeometric equation, where two the branch points of the hypergeometric equation merge at $\infty$ to become an essential singularity. The origin is the regular singularity. The two parameters of the equation determine the Regge trajectories, and the degeneracy of the angular momentum states. The Regge trajectories express the angular momentum in terms of the energy, and if the potential is real, or the energy is greater than the potential energy,  the angular momentum will be real and discrete. The parameter that determines the trajectories discriminates between quantized motion of bound states and unstable resonance states. In the former case it is negative, and identified as the radial quantum number, whereas in the latter case it can be associated with a \lq complex\rq\ angle.  In the case of the nonrelativistic coulomb interaction,  the Bohr formula for the energy levels of the hydrogen atom result when the parameter is set equal to a negative integer, which is also the index of the Laguerre polynomials that represent the radial component of the wave function. All this can be obtained without the usual procedure of expanding the radial wave function in a series and terminating it at a certain point to obtain a quantum condition. 

Rather, if the energy becomes positive,  the potential complex, or the potential energy  greater than the total energy,  quantization does not occur, and the parameter represents an angular point whose homologue is branch point. The new point is the generalization of the angular point to complex values. This is not unlike the generalization of the scattering amplitude to make it a function of the angular momentum. In order to make the scattering amplitude a function of the angular momentum, it had to be made complex and continuous so that one could take advantage of Poincar\'e's theorem which says that if a parameter in a differential equation, the angular momentum, or the wavenumber, appears only in the analytic function in some domain of the parameter, and if some other domain a solution of the equation is defined by a boundary condition that is independent of the parameter, then this solution is analytic in the parameter in the domain formed from the intersection of the two domains. In other words, Poincar\'e had shown that, under suitable conditions, the smooth solution to the differential equation could be made an analytic function of the parameters of that equation by allowing them to become complex and continuous, instead of real and discrete. 

The reason for naming the trajectories after Regge\cite{Regge} was that he  brought Poincar\'e's theorem to the attention of high energy physicists. Instead of confining his attention to integer angular momenta, Regge transformed the scattering amplitude so that it became a function of the, continuous, angular momenta. In order to do so, he had to allow it to become complex.  Regge's idea was not new, it had already been used by Poincar\'e himself, and Nicholson in 1910, to describe the bending of electromagnetic waves by a sphere. Sommerfeld and Watson used it to describe the propagation of radio waves on the surface of the earth, and their scattering from various potentials. It has become known as the Sommerfeld-Watson representation.

The new, unphysical, regions provided proving grounds for speculative high energy physics. The passage from a real to a complex parameter is not nearly as radical as that from a discrete to a continuous one, or from a positive to a negative one. How does one define \lq negative\rq\ angular momenta? The angular momentum is represented in the equation for the energy as a centrifugal repulsion. At a constant attractive potential, the only way the energy could be made more negative, thereby allowing more bound states to be formed, is to convert a centrifugal repulsion into a centripetal \lq attraction\rq\ by allowing the angular momentum to become negative. The limit occurs where the two indicial solutions to the Schr\"odinger equation with a centrifugal term coincide. One is called the \lq regular\rq\ solution because it goes to zero at the origin, while the other is the \lq irregular\rq\ solution because it blows up there. Conventionally, the latter solution is rejected because any admixture of the two would not lead to a unique solution. However this is incorrect because \emph{both\/} indicial exponents determine how the ratio of the solutions, which is an automorphic function, transform. 

An automorphic function is a periodic function under the group of linear (fractional) substitutions. When Poincar\'e came on the scene in 1880 the only two periodic functions that were known were the trigonometrical and elliptical functions. By cutting and pasting edges of the fundamental region together one could get solid figures with different amount of holes, or genus. Trigonometric functions had no holes, elliptic functions, one hole belonging to a torus, and Poincar\'e wondered if there were automorphic functions with a greater number of holes. Any given point of the fundamental region would be transformed into the same point in an adjacent fundamental region by the linear fractional transformation. It would not connect points in the same fundamental region, for, otherwise, it would not be \lq fundamental\rq.

For instance, if the angular momentum is negative and in the interval $[0,-\half]$, the plane would be tessellated by crescents formed from the intersection of nonconcentric circles whose angle would be the degeneracy of states. It is precisely in the unphysical region that the greater than unity cosine has become a hyperbolic cosine with a complex angle. The crescent of the plane with a given angle will be successively transformed by the fractional linear transformation ultimately returning to itself. Thus, the entire plane is divided into portions equal in number of the periodic order of the substitution.

When two particle collide there is a scattering angle, $\vartheta$, whose cosine  resides between $-1$ and $+1$ in the physical region. However, by allowing $\cos\vartheta$ to go to either plus or minus infinity, enables one to consider infinite momentum transfer. Such large momentum transfer occur over extremely small distances. Thus, large $\cos\vartheta$ is adapted to the study of strongly scattered waves that can bind and resonate. 

Poles can be expanded in a bilinear series of the products of Legendre functions of the first and second kinds. Only when the pole is within a given ellipse will the series converge. The ellipse is determined by the trace of the monodromy matrix which is a hyperbolic cosine with a complex argument. The hyperbolic cosine of the real component is the semi-major axis of the ellipse, while the imaginary component is the eccentric angle of the ellipse. In the case of coulomb scattering the real component is the ratio of the charges to the velocity of the incoming particle so that the ellipse will be larger the smaller the velocity of the incoming particle. This is referred to as the classical region. As the velocity increases we are transformed into the relativistic region with a decrease in the size of the ellipse.

What is a complex angle? In optics, complex angles arise when a refractive wave does not penetrate into the second medium, but, rather, propagates parallel to the surface. The system is then said to suffer total internal reflection. There is no energy flow across the surface, and at the angle of incidence there is total reflection. For angles of incidence greater than the critical value, the angle of reflection becomes complex with a pure imaginary cosine meaning the wave is attenuated exponentially beyond the interface. 

The three poles of the second order differential equation are associated with the thresholds of particle creation in high energy physics.\cite{Eden} Their \lq residues\rq\ are given by the angles of a triangle which tessellate the complex plane. The angles themselves may be complex like the argument of the hyperbolic cosine above. In order to tessellate the plane it must reproduce itself by rotating about any given axis. When the angles become complex, it will not only rotate the sides of the triangle but will also deform them. The fact that any two angles are complex conjugates, related to source and sink, will render the sum of the angles  real but, may not be equal to $\pi$. 

For Regge trajectories these angles represent the complex angular momentum. Below threshold the imaginary component of the angle momentum vanishes for there is no state to decay into so the resonance width is zero.
Regge gave an interesting interpretation to the imaginary component of the angular momentum. Just as the longer the time the smaller the uncertainty in energy, so that long time uncertainty is related to small resonance widths, imaginary angular momentum is related to change in angle through which the particle orbits during the course of a resonance. For extremely long resonances, the angle of orbit is large until it becomes permanent in a bound state.

Triangle functions which tessellate the complex plane are described by automorphic functions which represent solid figures. The  automorphic functions are the inverses of the quotients of the two independent solutions to a second order differential equation. These quotients can be moved around the complex plane by linear fractional transformations which delineate the fundamental regions. Schwarz showed how these automorphic functions map one complex plane onto another, just like the hyperbolic cosine maps ellipses in one plane onto circles in another.  The angles are either the interior, or exterior, angles of the triangle which is the fundamental region. In the case of large, but real, $\cos\vartheta$, the automorphic function is none other than the expression for the Legendre function of the second kind at a large value of its order, the angular momentum.\cite{BHL} It tessellates the surface of a sphere with triangles whose bases lie on the equator of the sphere, each angle being $\pi$ radians, and one vertex at the north pole whose angle is proportional to the difference between the order of the Legendre function and the angular momentum of the Regge trajectory. The solid is a double pyramid, or a dihedron, whose triangles have sums greater than $\pi$ radians, and, therefore, belong to elliptic geometry.

In fact, this automorphic function has been proposed as a partial wave scattering amplitude. Another proposal was made by Veneziano\cite{Veneziano} who showed that the Euler beta integral satisfies the duality principle of high energy physical where the scattering amplitude remains the same under the exchange of total energy and momentum transfer. Experimentally, this is achieved by replacing the particle with its anti-particle.  It so happens that the beta integral is the automorphic function of Schwarz for triangle tessellations. The angles are the Regge trajectories which become complex above threshold. In fact, a beta integral with real arguments could not represent a complex scattering amplitude for it would be physically measureable being the distance between the angles in the triangle. The Veneziano model, which has served as the impetus of string theories,  was found to be wanting in the hard sphere limit because it did not reflect the granular, or parton-like, behavior observed in deep inelastic scattering experiments.

What is the use of automorphic functions in high energy particle physics? First, it provides restrictions on the nature of the complex Regge trajectories and on the nature of the potentials. The potentials must be real for bound states, complex, or imaginary, for resonances. Second, the possibility of their being a complementarity between continuous groups in quantized systems and discrete groups with a continuous range of non-quantized parameters for bound states and resonances. Third, the spectrum of resonances that lie along a Regge trajectory is likened to the nesting and proliferation of circles when the number of generators is increased.

\section{Nonrelativistic Coulomb Interaction}
 The confluent hypergeometric equation arises from the confluence of two singularities in Riemann's hypergeometric equation leaving the regular and irregular singularities at $0$ and $\infty$. It can therefore describe an infinite-range potential like the coulomb potential for which it is given by~\protect\footnote{This is formerly identical with Eqn. (26) on page 52 in Ref.~\onlinecite{Mott}. However, the definition of $\eta$ there is real. The same expression can be found in Ref.~\onlinecite{Gottfried}. But then the quantization condition $a=-n^{\prime}$ is complex.\cite[Eq.~(27) p. 156]{Gottfried} Since  $\sqrt{2E}$  is imaginary for bounded states, and not the velocity $v$, the quantization condition is real. Rather, in Ref.~\onlinecite{Singh} the condition $a=-n^{\prime}$ is taken as the condition of the poles of the $S$-matrix, giving the position of the $n^{\prime}$ Regge pole.} 
 \begin{equation}
\frac{d^2\psi}{dr^2}+\left(\frac{c}{r}-1\right)\frac{d\psi}{dr}-\frac{a}{r}\psi=0. \label{Kummer}
\end{equation}
The parameters for the coulomb interaction are 
\begin{equation}
c=2(\ell+1),\label{c}
\end{equation}
 and 
 \begin{equation}
 a=\ell+1-i\eta, \label{a}
 \end{equation}
where $\ell$ is the total angular momentum, $\eta=ZZ^{\prime}e^2/\surd(2E)$ is the coulomb parameter   which is negative if the charges $Ze$ and $Z^{\prime}e$ are opposite, and $E$ is the energy of the incoming particle in units $\hbar=c=m=1$. 

The parameter $a$ determines the nature of the trajectories. If $a=-n^{\prime}$, the system has bound states where $n^{\prime}$ is the radial quantum number, and the total energy, $E<0$. This specifies the Kummer function $\psi$ as a Laguerre polynomial of index $n^{\prime}$. Specifying the Regge trajectory (\ref{a}) avoids the introduction of a series expansion  in the Schr\"odinger equation, and imposing a cut-off.   When $\eta=0$, the solution to Eq.~(\ref{Kummer}) reduces to a product of an exponential function and a hyperbolic Bessel function.

The confluent hypergeometric equation, (\ref{Kummer}), can be easily converted into the Schr\"odinger equation,
\begin{equation}
\frac{d^2\psi}{dr^2}+\fourth\left(\frac{1-\lambda^2}{r^2}+4i\frac{\eta}{r}-1\right)\psi=0,\label{Schrodinger}
\end{equation}
by the substitution $\psi\rightarrow e^{-\half\int^{r}(c/r-1)dr}\psi$, where $\lambda=2\ell+1$. Alternatively, if we didn't know Eq.~(\ref{Kummer}), we could transform (\ref{Schrodinger}) into it by inverting the substitution. The indicial equation as $r\rightarrow0$ is
\begin{equation}
\psi=Ar^{-\ell}+Br^{\ell+1},\label{indices}
\end{equation}
where the constant $A$ is conventionally set equal to zero in order for $\psi$ not to diverge at the origin. However, at $\ell=-\half$ both terms give the same dependence upon $r$. It is precisely at $\ell=-\half$ where the regular, $B$, and irregular, $A$, solutions in Eq.~(\ref{indices}), coincide. There is no reason to constrain the angular momentum to positive integral or semi-integral values since we are considering \lq elementary\rq\ and \lq composite\rq\ particles, which may be stable or unstable.

We may look for a solution to Eq.~(\ref{Kummer}) in the form of a Laplace transform~\cite{Gottfried},
\begin{equation}
\psi(r)=\int_{\kappa_{1}}^{\kappa_{2}}e^{\kappa r}\phi(\kappa)\;d\kappa \label{Laplace},
\end{equation}
 for the (normalized) wavenumber, $\kappa$. Introducing it into (\ref{Kummer}) results in
\begin{center}
 \begin{eqnarray*}
\int_{\kappa_{1}}^{\kappa_{2}}\phi(\kappa)\left\{(c\kappa-a)+\kappa(\kappa-1)\frac{d}{d\kappa}\right\}e^{\kappa r}d\kappa\\
  =\phi(\kappa)\kappa(\kappa-1)e^{\kappa r}
\Bigg|_{\kappa_{1}}^{\kappa_{2}}+\int_{\kappa_{1}}^{\kappa_{2}}\left[(c\kappa-a)-\frac{d}{d\kappa}\kappa(\kappa-1)\right]\phi(\kappa)\;d\kappa,
 \end{eqnarray*}
\end{center}
 after an integration by parts has been performed.
 If the limits $\kappa_1$ and $\kappa_2$ can be chosen so as to make the integrated part to vanish, then $\phi(\kappa)$ will satisfy
 \[
 \left[(c\kappa-a)-\frac{d}{d\kappa}\kappa(\kappa-1)\right]\phi(\kappa)=0. \]
 The solution to this first order equation is
 \begin{equation}
 \phi(\kappa)=C\kappa^{a-1}(1-\kappa)^{c-a-1},\label{phi}
 \end{equation}
where $C$ is a constant of integration.
 In view of the Laplace transform, (\ref{Laplace}), we find
 \begin{equation}
 \psi(r)=C\int_{\mathcal{C}}\kappa^{a-1}(1-\kappa)^{c-a-1}e^{\kappa r}\;d\kappa, \label{psi}
 \end{equation}
 where, if the contour $\mathcal{C}$ is not closed the integrand in (\ref{psi}) is required to have the same value at the endpoints. 
 
 In contrast to the original confluent hypergeometric equation (\ref{Kummer}), which has branch points at $0$ and $\infty$, we have added an additional branch point at $1$ by considering the wave number. This branch point may be thought of as placing a bound,
\[
p\le \surd(2E), \]
on the  momentum $p$ by the square root of twice the total energy, like the maximum momentum of a Fermi gas of elementary particles at absolute zero.\cite{LL}

The difference between (\ref{c}) and (\ref{a}) determines the second angle as
\begin{equation}
(c-a)\pi=\left(\ell+1-i\eta\right)\pi=\bar{a}\pi, \label{c-a}
\end{equation}
which is the complex conjugate of (\ref{a}). We will appreciate that attractive and repulsive coulomb potentials always appear as complex conjugates, or, equivalently, for every source there is a sink.

If $y_1$ and $y_2$ are any two particular solutions to the hypergeometric equation, then the Wronskian is given by~\cite{Forsyth}
\[
y_2\frac{dy_1}{dr}-y_1\frac{dy_2}{dr}=Cr^{-c}(1-r)^{c-a-b-1},\]
where $C$ is a constant.
Dividing both sides by $y_2^2$, it becomes the derivative of the ratio $w=y_1/y_2$ of the two particular solutions. Now, any other two solutions, say $y_1^{\prime}$ and $y_2^{\prime}$ can be expressed as linear combinations of $y_1$ and $y_2$, viz.,
\begin{eqnarray*}
y_1^{\prime} &=&\alpha y_1+\beta y_2\\
y_2^{\prime} &=&\gamma y_1+\delta y_2
\end{eqnarray*}
so that the quotient of the new solutions, $w^{\prime}=y_1^{\prime}/y_2^{\prime}$ is related to the quotient of the old, $w$, by a linear fractional transformation 
\begin{equation}
w^{\prime}=\frac{\alpha w+\beta}{\gamma w+\delta}. \label{Mobius}
\end{equation}
When the quotient of the solutions is inverted, we get  a function automorphic with respect to a certain group of the linear fractional transformations. For the hypergeometric equation, the group of automorphisms are triangular tessellations of the unit disc. 

Our interest will be focused on the momentum space at $r=0$ in (\ref{Laplace}). Since the hypergeometric equation with the coefficient $b=0$,
\begin{equation}
\frac{d^2Y_2}{d\kappa^2}+\left(\frac{1-a}{\kappa}-\frac{1-\abar}{1-\kappa}\right)\frac{dY_2}{d\kappa}=0, \label{Kummer-bis}
\end{equation}  has one solution,
\begin{equation}
Y_2(a,c,\kappa)=C\int_0^{\kappa}t^{a-1}(1-t)^{\abar-1}dt, \label{Beta}
\end{equation}
which is an incomplete beta function, $B(a,\abar;\kappa)$ if the constant of integration is set equal to unity. The second solution
\begin{equation}
Y_1=w\;Y_2,\label{w-ratio}
\end{equation}
is given by an automorphic function $w^{-1}(\kappa)$. 

Dividing the Wronskian,
\[
Y_2\frac{dY_1}{d\kappa}-Y_1\frac{dY_2}{d\kappa}=C\kappa^{a-1}(1-\kappa)^{c-a-1},\]
 by $Y_2^2$, it becomes the derivative of the ratio, 
viz.,
\[w^{\prime}=C\frac{\kappa^{a-1}(1-\kappa)^{\abar-1}}{Y_2^2},\]
which has the Schwarzian derivative,
\begin{eqnarray}
\{w,\kappa\}& :=&\frac{w^{\prime\prime\prime}}{w^{\prime}}-\threehalves\left(\frac{w^{\prime\prime}}{w^{\prime}}\right)^2\nonumber\\
&=&\frac{1-a^2}{2\kappa^2}+\frac{1-\abar^2}{2(1-\kappa)^2}-\frac{(1-a)(1-\abar)}{\kappa(\kappa-1)}, \label{Schwarzian}
\end{eqnarray} 
 where the prime now stands for the derivative with respect to $\kappa$.

 With the transformation, 
 \[Y=\tilde{Y}e^{c\ln\kappa+(\abar-1)\ln(1-\kappa)}, \]
(\ref{Kummer-bis}) can be converted into
\begin{equation}\tilde{Y}^{\prime\prime}+I\tilde{Y}=0,\label{de}
\end{equation}
where $2I=\{w,\kappa\}$, with the Schwarzian derivative given by (\ref{Schwarzian}). The Schwarzian derivative has a long and glorius history dating back to Lagrange's investigations on stereographic projection used in map making.~\cite{Ovsienko}

The third angle can be read off from the Schwarzian, (\ref{Schwarzian}), and is
\begin{equation}
(c-1)\pi=\lambda\pi, \label{c-1}
\end{equation}
which is the original angle of the crescent, having branch points at $0$ and $\infty$. This is due to the centrifugal potential in the Schr\"odinger equation. The coulomb potential introduces the complex conjugate angles, $a$ and $\abar$, which make the interaction independent of whether it is attractive or repulsive since they appear symmetrically. If we adhere to the triangle representation, the requirement that the angles be less than $\pi$ limits $\ell$ to the closed interval $[-\half,0]$. In this interval,  centrifugal repulsion $\ell(\ell+1)/r^2$ becomes centrifugal \lq attraction.\rq\ 

For $\ell=-\half$, the sum of the \lq angles\rq\ of the triangle, $(\half-i\eta)\pi$, $(\half+i\eta)\pi$, and $0$ is $\pi$.\cite{Choudhary} The analytic function $B(a,\abar;\kappa)$ given by (\ref{Beta}) for $C=1$ maps the upper half-plane $\Im\{\kappa\}>0$ onto the interior of a \lq half-strip\rq\ formed by the two base \lq angles\rq\ $a\pi$ and $\abar\pi$ corresponding to the points $\kappa=0$ and $\kappa=1$ in the $\kappa$-plane. The distance between the two vertices in the $B$-plane is
\begin{equation}
\int_0^1 t^{a-1}(1-t)^{\abar-1}dt=\Gamma\left(\half-i\eta\right)\;\Gamma\left(\half+i\eta\right)=\frac{\pi}{\cosh(\eta\pi)}
\label{Beta-bis}
\end{equation}

In general, the amplitude will be given by the complete Beta function
\begin{equation}
B(a,\abar;1)=\frac{\Gamma(\ell+1-i\eta)\;\Gamma(\ell+1+i\eta)}{\Gamma(2(\ell+1))}.\label{beta-tris}
\end{equation}
The beta function is symmetric with respect to attractive and repulsive potentials. In the attractive case, there is an infinite number of Reggie poles determined by the poles in the numerator of (\ref{beta-tris})\cite{Singh}, i.e.,
\begin{equation}
\ell+1-\frac{ZZ^{\prime}e^2}{\sqrt{2|E|}}=-n^{\prime}. \label{bound-Regge}
\end{equation}
These are bound states with $E<0$. In the case of resonances, $E>0$, the imaginary parts of the numerator of (\ref{beta-tris}) are equal and opposite so as to preserve the analyticity of the amplitude. In the case of a repulsive potential, with $E<0$, Eq.~(\ref{beta-tris}) tells us there there will again be an infinity of Regge poles. However, these will not correspond to bound states or resonances because the Regge poles are restricted to the left half of the $\ell$ plane. This information is contained in the amplitude, and it is equivalent to two $S$-matrix elements, one being the inverse of the other.\cite{Singh} 

Parenthetically, we would like to point out that the Regge pole behavior for an infinite-range potential is quite different than that for short range ones. A Regge trajectory for a short-range potential would have the form
\begin{equation}
\ell(E)=-1\pm \ell^{\prime}(E)\sqrt{-E}, \label{Regge-short}
\end{equation}
where the prime stands for differentiation. The intercept of $-1$ is called for by the form of the angular momentum term $\ell(\ell+1)$ in the Schr\"odinger equation. The fact that the cross-sections do not vanish asymptotically with the energy, but increase slowly with it is attributed to an exchange of a Reggeon whose intercept is $+1$.\cite{Forshaw} However, a positive intercept cannot be interpreted as originating in the angular momentum. This is substantiated by the fact that only negative intercepts give rise to the conservation of angular momentum in hyperbolic space.\cite{Lavenda} 

In Eq.~(\ref{Regge-short}) there would be no violent jump in the trajectories as  the energy passes through zero. In the case $E>0$ the poles are complex, but there is no violation of analyticity since they are complex conjugates. In the asymptotic case of large angular momentum the amplitude (\ref{beta-tris}) will be modulated by oscillations, viz.,
\[
B(a,\abar;1)\stackrel{\ell\gg1}{\longrightarrow}e^{\eta^{2}}\left(\frac{\ell+i\eta}{\ell-i\eta}\right)^{i\eta}.\]

\begin{figure}[htbp]
	\centering
		\includegraphics[width=0.35\textwidth]{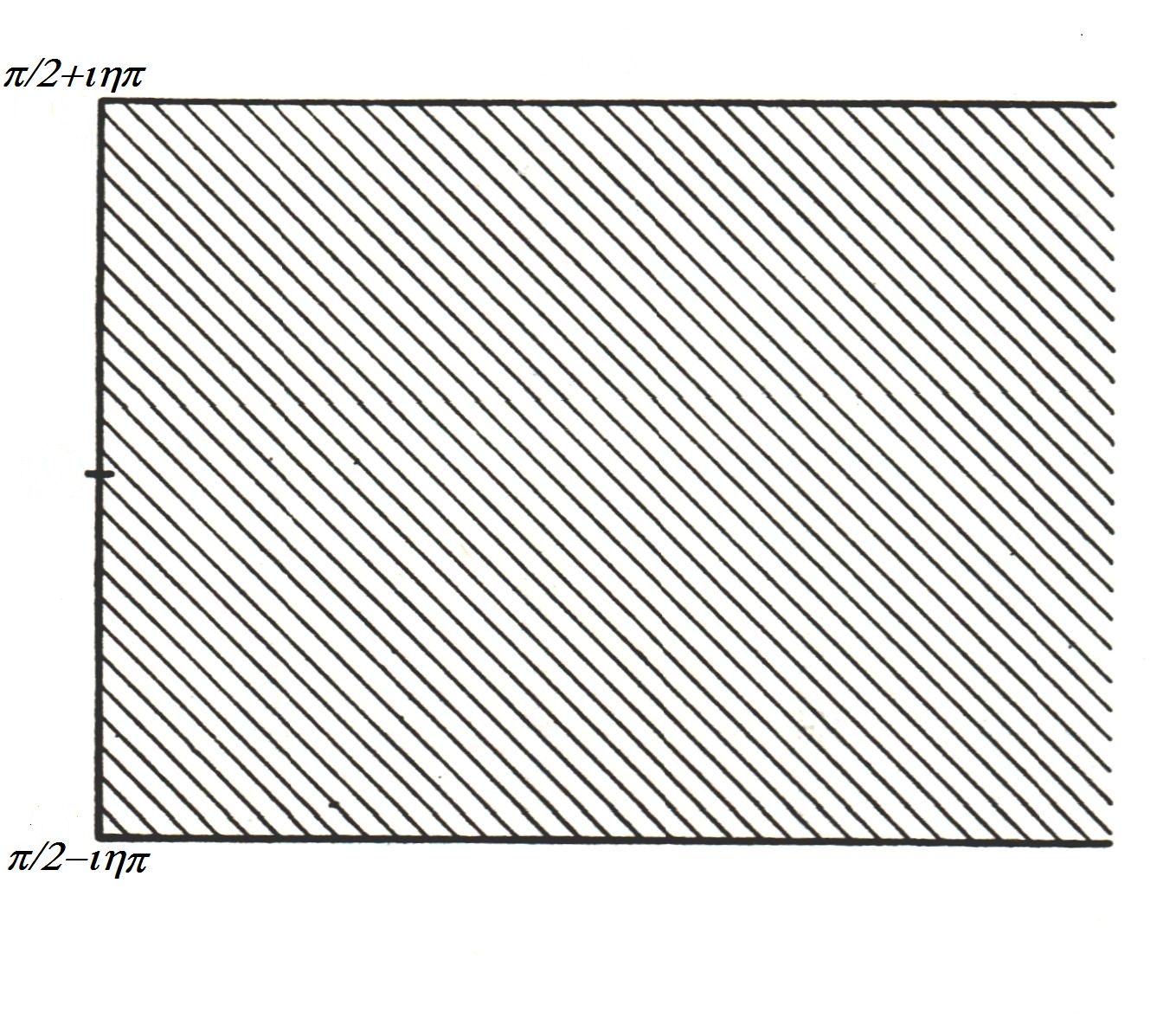}
	\caption{The half-strip obtained by the conformal mapping Eq.~(\ref{Beta-tris}).}
	\label{strip}
\end{figure}

The Schwarz-Christoffel transform remains valid even when one of the vertices of the triangle coincides with the point at infinity. The lengths of the sides from either vertex to the vertex at infinity are infinite, so that \begin{equation}
B(a,\abar;\kappa)=\int_0^{\kappa}t^{-\half-i\eta}(1-t)^{-\half+i\eta}dt=\int_0^{\kappa}\left(\frac{1-t}{t}\right)^{i\eta}\frac{dt}{\sqrt{t(1-t)}} \label{Beta-tris}
\end{equation}
will map the upper half-plane $\Im\{\kappa\}>0$ onto the interior of the \lq half-strip\rq\ shown in FIG.~\ref{strip}.

\section{Generators from Monodromy Relations}
If 
\begin{equation}
\lim_{\kappa\rightarrow0}\kappa^2I=\rho,\label{lim}
\end{equation}
the indicial equation for the branch point $0$ is
\[n(n-1)+\rho=0.\]
The two unequal roots, $n_1$ and $n_2$,
 are the exponents of the integrals of the equation
\begin{eqnarray*}
Y_1&=&\kappa^{n_1}+\cdots\\
Y_2&=&\kappa^{n_2}+\cdots,
\end{eqnarray*}
so that ratio of the solutions is
\begin{equation}
w=Y_1/Y_2=\kappa^{n_1-n_2}+\cdots \label{omega-bis}
\end{equation}
Since the discriminant of the quadratic form is
\[1-4\rho=a^2,\]
we have
\[I=\fourth\frac{1-a^2}{\kappa^2}+\cdots\]
in view of (\ref{lim}).

The ratio of the two solutions, (\ref{omega-bis}), can be written as
\begin{equation}
w=\kappa^{\ell+1-i\eta}+\cdots\label{omega-tris}
\end{equation}
According to Riemann, the exponents of the indicial equation, 
\begin{equation}
n_{1,2}=\half(1\pm a), \label{indicial}
\end{equation}
\lq\lq completely determine the periodicity of the function\rq\rq\cite{Gray}, in reference to the solutions of the hypergeometric equation. Eq.~(\ref{indicial}) shows that the difference in roots of the indicial equation, or exponents as they are commonly referred to, is the angle at the singular point. The monodromy relations imply that the roots to the indicial equations are rational numbers, which, in turn imply elliptic generators. Here, they are generalized to complex numbers so as to allow for loxodromic generators, and take as a definition of monodromy as the invariance of automorphic functions, or functions inverse to the quotient of two independent solutions of a differential equation, under a certain group of transformations.

A circuit around $\kappa=\infty$ in the positive direction that returns $Y_1$ as $e^{i2\pi n_1}Y_1$, while it returns $Y_2$ as $e^{i2\pi n_2}Y_2$,  is performed by
\[
\mathbb{M}=\left(\begin{array}{cc} e^{2\pi i n_1} & 0\\ 0 & e^{2\pi i n_2}\end{array}\right).
\] The monodromy theorem asserts  that any global analytic function can be continued along all curves in a simply connected region that determines a single-valued analytic function on every sheet of the branch points, one sheet per branch point. The monodromy matrix about $\kappa=\infty$,
\begin{equation}
\mathbb{C}=\left(\begin{array}{cc}
e^{-2\pi i\ell} & 0\\ 0 & e^{2\pi i\ell} \end{array}\right),\label{C}
\end{equation}
has unit determinant, and trace
\[\mathsf{Tr}\;\mathbb{C}=2\cos(2\pi\ell),\]
so that it is an elliptic transformation. Rather, around $\kappa=0$, the  monodromy matrix,
\begin{equation}
\mathbb{A}=\left(\begin{array}{cc}
e^{i\pi(\ell+i\eta)} & 0\\ 0 & e^{-i\pi(\ell+i\eta)} \end{array}\right),\label{A}
\end{equation}
also has unit determinant, but trace,
\begin{equation}
\mathsf{Tr}\;\mathbb{A}=2\cos\left(\ell+i\eta\right)\pi, \label{Tr-A}
\end{equation}
so that it is a loxodromic generator. This is also true around $\kappa=1$, where
\begin{equation}
\mathbb{B}=\left(\begin{array}{cc}
e^{i\pi(\ell-i\eta)} & 0\\ 0 & e^{-i\pi(\ell-i\eta)} \end{array}\right),\label{B}
\end{equation}
 whose trace is
\begin{equation}
\mathsf{Tr}\;\mathbb{B}=2\cos\left(\ell-i\eta\right)\pi. \label{Tr-B}
\end{equation}
The product of any two matrices, or their inverses, will yield the third, or its inverse. Stated slightly  differently, a circuit around two of the branch points in a positive (anti-clockwise) direction will give a circuit around the third branch point in the negative (clockwise) direction, since $\mathbb{A}\mathbb{B}\mathbb{C}=\mathbb{I}$ is the unit matrix.

The matrices are abelian since they have the same fixed points, $0$ and $\infty$. $\mathbb{A}$ and $\mathbb{B}$ are loxodromic because their traces, (\ref{Tr-A}) and (\ref{Tr-B}), are complex and $>2$. Loxodromic transformations do not leave the disc invariant: they transfer the inside of the disc to the outside of its inverse.

The generators, (\ref{A}), (\ref{B}), and (\ref{C}), have the fixed points $0$ and $\infty$ in common.
Conjugation with the Cayley mapping,
\begin{equation}
\mathbb{K}(z)=\frac{z-i}{z+i}, \label{Cayley}
\end{equation}
carries these fixed points to $-1$ and $+1$, respectively, while its inverse, $\mathbb{K}^{-1}$, takes them to $i$ and $-i$, respectively. For instance, the conjugate generator  of (\ref{A}), 
\begin{equation}
\mathbb{W}=\mathbb{K}\mathbb{A}\mathbb{K}^{-1}=\left(\begin{array}{cc} \cosh\pi(\eta+i\ell) & \sinh\pi(\eta+i\ell)\\ \sinh\pi(\eta+i\ell) & \cosh\pi(\eta+i\ell) \end{array}\right),
\label{W}
\end{equation}
carries the fixed points $0$ and $\infty$ to $-1$ and $1$, respectively,
while those of the inverse conjugation,
\begin{equation}
\mathbb{T}=\mathbb{K}^{-1}\mathbb{A}\mathbb{K}\left(\begin{array}{cc} \cosh\pi(\eta+i\ell) & -i\sinh\pi(\eta+i\ell)\\ i\sinh\pi(\eta+i\ell) & \cosh\pi(\eta+i\ell) \end{array}\right),
\label{T}
\end{equation}
carry the fixed points to $i$ and $-i$, respectively.

The loxodromic transformation, (\ref{W}), pairs the isometric circle, $\mathfrak{I}_w$, with its inverse, $\mathfrak{I}_w^{\prime}$. That is, $\mathbb{W}$ carries points in the interior of $\mathfrak{I}_w$ to the exterior of $\mathfrak{I}_w^{\prime}$. The fixed points of (\ref{W}), $-1$ and $1$, lie in $\mathfrak{I}_w$ and $\mathfrak{I}_w^{\prime}$, respectively. With $\ell=0$, (\ref{W}) is pure stretching, pushing points from $-1$, the source, to $+1$, the sink. 

The cyclic group consists of one generator, and applying it $n$ times gives
\begin{equation}
\mathbb{W}^n= \left(\begin{array} {cc}\cosh\pi(n\eta) & \sinh\pi(n\eta)\\ \sinh\pi(n\eta) & \cosh\pi(n\eta) \end{array}\right),
\label{W-bis}
\end{equation}
for $\ell=0$. Whereas the isometric circle of (\ref{W}) is
\begin{equation}
\left|\sinh(\pi\eta)z+\cosh(\pi\eta)\right|=1, \label{cir}
\end{equation}
the isometric circle of (\ref{W-bis}) is
\begin{equation}
\left|\sinh(n\pi\eta)z+\cosh(n\pi\eta)\right|=1. \label{cir-bis}
\end{equation}
The isometric circle, (\ref{cir}), has its center at $-\coth(\pi\eta)$, and radius $R_1=1/\sinh(\pi\eta)$. The isometric circle, (\ref{cir-bis}), on the other hand, has its center at $-\coth(n\pi\eta)$, and radius $R_n=1/\sinh(n\pi\eta)$. Since $\sinh(n\pi\eta)>\sinh\pi\eta$, it follows that $R_1>R_n$. And since this is true for any $n>1$, the isometric circles will be nested inside one another, becoming ever smaller until the limit point is reached.

 The loxodromic generator, (\ref{T}), can be decomposed into a product, 
\begin{eqnarray*}\left(\begin{array}{cc}
\cosh\pi(\eta+i\ell) & -i\sinh\pi(\eta+i\ell)\\ i\sinh\pi(\eta+i\ell) & \cosh\pi(\eta+i\ell)\end{array}\right)
=\left(\begin{array}{cc}\cosh(\pi\eta) &-i\sinh(\pi\eta)\\ i\sinh(\pi\eta) & \cosh(\pi\eta)\end{array}\right)
\left(\begin{array}{cc} \cos(\pi\ell) & \sin(\pi\ell)\\ -\sin(\pi\ell) & \cos(\pi\ell) \end{array}\right)\\
=\mathbb{V}\;\mathbb{U}
\end{eqnarray*}
 of hyperbolic, $\mathbb{V}$, and elliptic, $\mathbb{U}$,  transformations, \emph{with the same fixed points\/}. The isometric circles of $\mathbb{V}$ and its inverse, $\mathbb{V}^{-1}$,  
 \[|\mp i\sinh(\pi\eta) z+\cosh(\pi\eta)|=1,\]
  have their centers on the imaginary axis at $z_0=\mp i\coth(\pi\eta)$, and radius $1/\sinh(\pi\eta)$. Part of the plane exterior to these two circles is the fundamental region for the group generated by $\mathbb{T}$.\cite{Ford} In other words, it does not depend on the elliptic transformation $\mathbb{U}$.

Since the isometric circle is defined by
\[\mathbb{V}^{\prime}=\left|-i\sinh(\pi\eta)z+\cosh(\pi\eta)\right|^{-2}=1,\]
and that of its inverse by
\[\mathbb{V}^{-1\;\prime}=\left|i\sinh(\pi\eta)\mathbb{V}(z)+\cosh(\pi\eta)\right|^{-2}=\left|-i\sinh(\pi\eta)z+\cosh(\pi\eta)\right|^2,\]
whatever is inside the isometric circle of $\mathbb{V}$, 
\[\left|-i\sinh(\pi\eta)z+\cosh(\pi\eta)\right|<1, \]
i.e., $\mathbb{V}^{\prime}>1$ is outside of its inverse, because $\mathbb{V}^{-1\;\prime}<1$, and vice versa. 

Let $\mathbb{W}$ and $\mathbb{V}$ be the generators which pair off the isometric circles, $\mathfrak{I}_w$ with its inverse, $\mathfrak{I}^{\prime}_w$, and $\mathfrak{I}_v$ with $\mathfrak{I}_v^{\prime}$, respectively. The matrices $\mathbb{W}$ and $\mathbb{V}$ will be external to one another provided $\cosh(\pi\eta)>\surd2$.~\protect\footnote{If the inequality becomes an equality, it is treaded by the example~Ref.~\onlinecite{Mumford} which is the condition that the four circles are tangent to one another. The trace of the commutator is $-2$ indicating that the two fixed points have coalesced into one at the point where the circles touch. Both groups are Fuchsian since their limit points are either on a line or on a circle.}
The matrix, (\ref{W}), has fixed points $\mp1$, and pairs circles with centers, $\mp\mbox{coth}(\pi\eta)$, and the same radii, $\/\sinh(\pi\eta)$, on the real axis. The fixed point $-1$ will be located in the circle on the negative  axis, while the attracting fixed point $+1$, which is a sink, will be located in the circle on the positive axis, since $\tanh\half(\pi\eta)<1$ for whatever value $\eta$ happens to be. 

Furthermore, let $\mathbb{Z}=\mathbb{V}\mathbb{W}$ be associated with the isometric circle $\mathfrak{I}_z$. If $\mathfrak{I}_v$ and $\mathfrak{I}_w^{\prime}$ are external to one another then $\mathfrak{I}_z$ is in $\mathfrak{I}_w$\cite[p. 53, Thm 12]{Ford}. For suppose that the circles are not tangent to one another, then if $p$ is a point outside of, and not on, $\mathfrak{I}_v$, the generator $\mathbb{V}$ will carry the point $p$ into, or on, $\mathfrak{I}_v^{\prime}$, say $p^{\prime}$ with a decrease in length, or at least no change in length. Since $p^{\prime}$ is outside of $\mathfrak{I}_w$, $\mathbb{W}$ will transform it with a decrease in length. Consequently,the combined operation, $\mathbb{Z}$, will transform $p$ with a decrease in length, implying the $p$ is outside of $\mathbb{I}_z$. And since every point on or \emph{outside\/} of $\mathfrak{I}_v$ is also outside of $\mathfrak{I}_z$, the latter must be inside the former. 

\begin{figure}[htbp]
	\centering
		\includegraphics[width=0.40\textwidth]{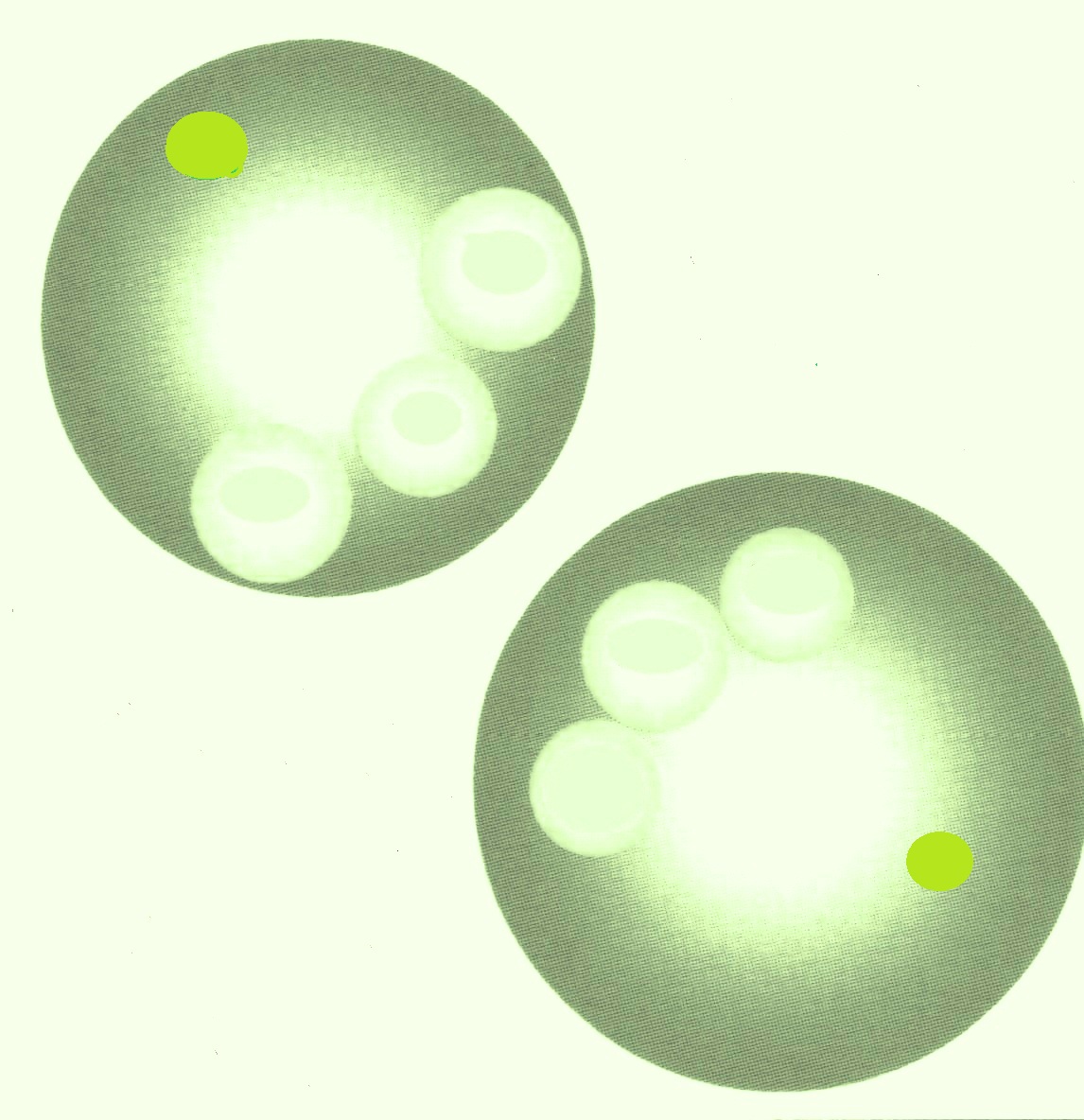}
	\caption{The nesting of circles in the isometric circle and its inverse.}
	\label{circles}
\end{figure}

Each time we add a generator, we get a nesting in a nesting of circles with their proliferation~\cite[p. 170]{Mumford}. The isometric circle $\mathfrak{I}_v$ will contain three nested circles, $\mathfrak{I}_z$, and $\mathfrak{I}_{z^{\prime}}$ for the generator $\mathbb{Z}^{\prime}=\mathbb{V}\mathbb{W}^{\prime}$, and another for $\mathbb{V}\mathbb{V}$. This is shown in FIG.~\ref{circles}.
 Each of the other three discs will also have three nested discs. Increasing the generator by one, so that there are now three generators, or \lq letters,\rq\ there will be three nested discs in each of the former discs, and so on. Thus, there would be no limit of an elementary particle, but, rather, particles within particles within particles and so on. There may result in high energy collisions additional particles to those of the compound particle disintegrating into its component parts because there may be sufficient energy that can be converted into matter before disintegrating again into other forms of matter. Moreover, the relativistic phenomenon of pair creation may be related to the pairs of isometric circles are considered, one for the generator of the transformation, and one for its inverse.

\section{The  Coulomb Phase Shift as a Projective Invariant}
The absolute conic for elliptic geometry is the null conic. It is defined in projective coordinates by an equation with real coefficients, but it is composed exclusively of imaginary points. The secant through the points  $k_1$ and $k_2$  join the conic at $-i$ and $i$. The cross ratio is
\begin{eqnarray}
(k_1,k_2;-i,i)&= &\frac{(k_1+i)(i-k_2)}{(k_2+i)(i-k_1)}=\frac{k_1k_2+1+i(k_2-k_1)}{k_1k_2+1-i(k_2-k_1)}\nonumber\\
&=&\frac{1+i\tan\varphi}{1-i\tan\varphi}=e^{2i\varphi},\label{Laguerre}
\end{eqnarray} 
where $\tan\varphi=\tan(\varphi_2-\varphi_1)$, and $k_i=\tan\varphi_i$, for $i=1,2$.

Equivalently, we can consider a real conic with imaginary points, $ik_1$ and $ik_2$, so that their join cuts the absolute at $-1$ and $1$. The cross ratio is now
 \begin{eqnarray}
(ik_1,ik_2;-1,1)&=&\frac{(ik_1+1)(1-ik_2)}{(ik_2+1)(1-ik_1)}=\frac{k_1k_2+1-i(k_2-k_1)}{k_1k_2+1+i(k_2-k_1)}\nonumber\\
&=&\frac{1-i\tan\varphi}{1+i\tan\varphi}=e^{-2i\varphi},\label{Laguerre-bis}
\end{eqnarray}
which is the complex conjugate of (\ref{Laguerre}). Thus, the euclidean angle, 
\begin{equation}
\varphi=\frac{1}{2i}\ln(k_1,k_2;-i,i)=-\frac{1}{2i}\ln(ik_1,ik_2;-1,1), \label{varphi}
\end{equation}
can be expressed in terms of a cross ratio, and, hence,  is a projective invariant. Thus, the logarithms of the cross ratios, (\ref{Laguerre}) and (\ref{Laguerre-bis}), are pure imaginary and many valued quantities. Thus, in order to associate a segment $k_1k_2$ with the logarithm of the cross ratio, a pure imaginary absolute constant must be chosen.\cite{Efimov}

\begin{figure}[htbp]
	\centering
		\includegraphics[width=0.30\textwidth]{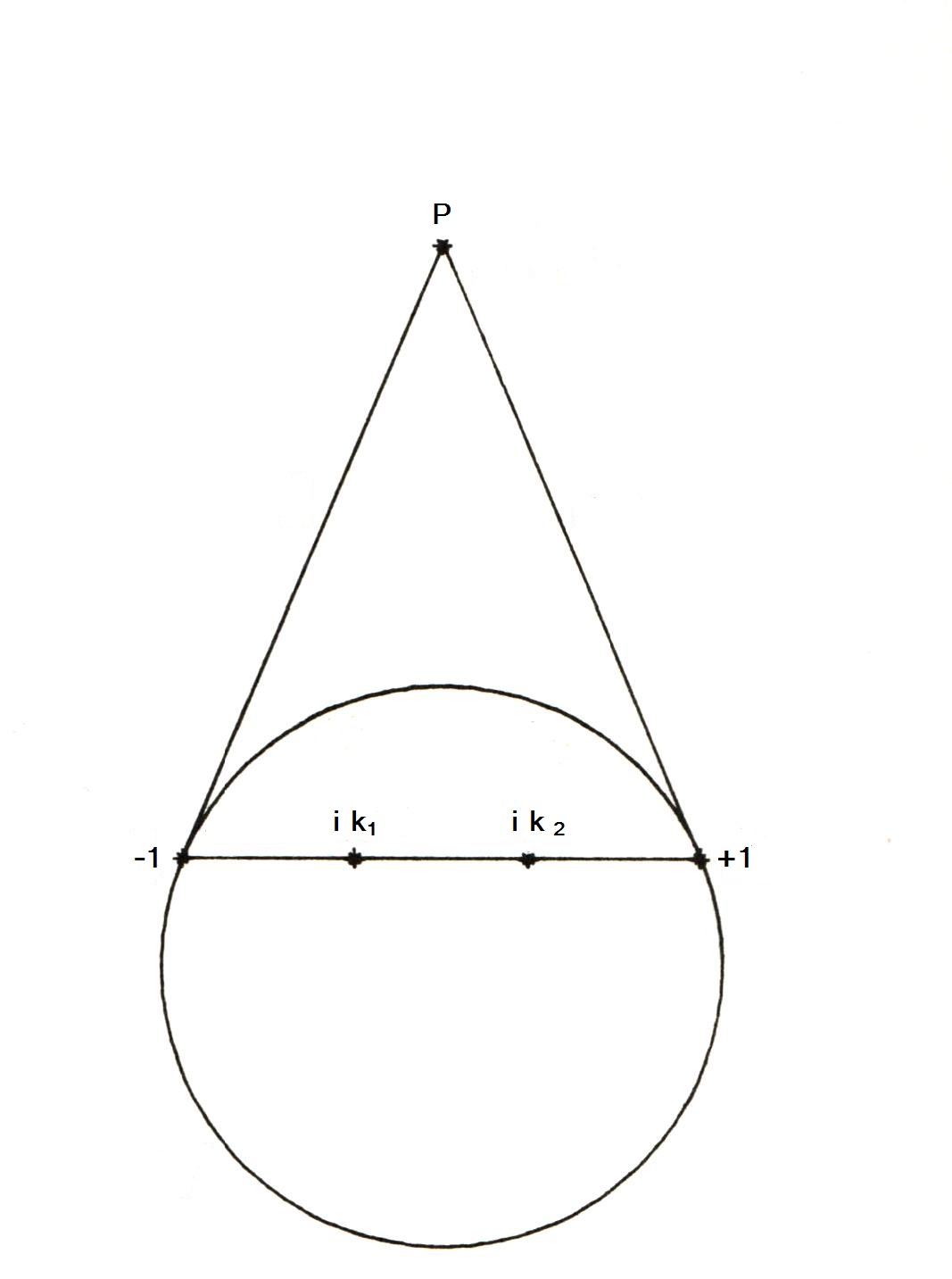}
	\caption{The real conic, tangents, pole $P$, and polar.}
	\label{pole}
\end{figure}

Conjugacy with respect to a polarity generalizes perpendicularity with respect to an inner product thus allowing euclidean geometry to be defined from affine geometry by singling out a polarity.\cite{Busemann} The imaginary points, $ik_1$ and $ik_2$, lie on the polar whose pole, $P$, is determined by the point of contact of two tangent lines to the real conic at points $-1$ and $+1$, as shown in FIG.~\ref{pole}.

The coulomb phase shift, $\sigma_{\ell}$, which determines purely electrostatic, or Rutherford, scattering is also a projective invariant. It  is defined by\cite{Blatt}
\begin{equation}
e^{2i\sigma_{\ell}}=\frac{(\ell+i\eta)!}{(\ell-i\eta)!}=\prod_{j=0}^{j=\ell-1}\frac{\ell-j+i\eta}{\ell-j-i\eta}e^{2i\sigma_0}. \label{ps}
\end{equation}
Transposing and taking the logarithm of both sides give
\begin{equation}
i\left(\sigma_{\ell}-\sigma_0\right)=\half\sum_{j=0}^{j=\ell}\ln\frac{\ell-j+i\eta}{\ell-j-i\eta}=\sum_{j=0}^{j=\ell-1}\tanh^{-1}\left(\frac{i\eta}{\ell-j}\right). \label{ps-bis}
\end{equation}
The coulomb phase shift is thus given by
\begin{equation}
\Delta \sigma=\sigma_{\ell}-\sigma_0=\sum_{j=0}^{j=\ell-1}\tan^{-1}\left(\frac{\eta}{\ell-j}\right), \label{BW}
\end{equation}
which is none other than a generalized Breit-Wigner expression\cite{Omnes}  in the neighborhood of a resonance where the angular momentum, $j$, stands in the for resonance energy, and $2\eta$ is the width.

It was Laguerre's great achievement to define euclidean geometry from affine geometry by singling out a polarity. The projective invariant (\ref{BW}) is a projective invariant in that it expresses an euclidean angle directly as the logarithm of the cross ratio.\cite{Busemann} It seems odd that the same name, Laguerre, should be associated with both the orthogonal polynomials when $a$ is a negative integer in the coulomb interaction and the derivation of euclidean geometry from affine geometry through a projective invariant.

\section{Relativistic Coulomb Interaction}
The relativistic generalization of the nonrelativistic coulomb interaction is given in this section. By specifying the coefficients in the confluent hypergeometric equation, we can obtain Dirac's expression for the energy of the hydrogen atom from the Klein-Gordon equation instead of the Dirac equation.

The coefficients in the confluent hypergeometric equation for the relativistic coulomb interaction are
\begin{eqnarray}
a&=&\half+\surd\left[(\ell+\half)^2-\gamma^2\right]+i\gamma E/\surd\left(E^2-m^2\right)\label{a-bis}\\
c&=&2\left(\half+\surd\left[(\ell+\half)^2-\gamma^2\right]\right),\label{c-bis}
\end{eqnarray}
where $\gamma=Z\alpha$,  $\alpha$ is the fine structure constant, and we have reinstated the mass $m$. If $a>0$ and $E>m$, we have scattering, and the second angle will be
\begin{equation}
c-a=\half+\surd\left[(\ell+\half)^2-\gamma^2\right]-i\gamma E/\surd\left(E^2-m^2\right). \label{c-a-bis}
\end{equation}
The two base angles, corresponding to the branch points at $0$ and $1$ in the $\kappa$-plane,  are, again, complex conjugates, the $+$ and $-$ signs, before the energy term  (\ref{a-bis}) and (\ref{c-a-bis}), correspond, respectively, to an attractive and repulsive coulomb potential.

The third angle is
\begin{equation}
(c-1)\pi=2\pi\surd\left[(\ell+\half)^2-\gamma^2\right], \label{s}
\end{equation}
and in order for it to be $<\pi$, the second inequality in
\begin{equation}
\gamma^2-\fourth\le \ell(\ell+1)\le \gamma^2, \label{ineq}
\end{equation}
has to be satisfied.
The other inequality is the condition that (\ref{s}) is real. The  upper bound converts a repulsive centrifugal forces into an attractive force, as can be seen in the Klein-Gordon equation, (\ref{KG}), below.

Alternatively, for $a<0$, and $E<m$, the Regge trajectories for the bound states are given by
\begin{equation}
-a=n^{\prime}=n-(\ell+1),\label{principal}
\end{equation}
where $n$ is the principal quantum number. This avoids  the necessity of looking for a series solution to the radial wave equation, and imposing a cut-off on the series. The condition that $a$ be equal to a negative integer implies that the Kummer function becomes a generalized Laguerre polynomial.  

Eq.~(\ref{principal}) gives the exact energy levels of a Dirac particle bound by a coulomb potential
\begin{equation}
E_{n,\ell}=m
\left[1-\frac{\gamma^2}{n^2+2(n-(\ell+\half))\left[\surd\left\{(\ell+\half)^2-\gamma^2\right\}-(\ell+\half)\right]}\right]^{\half}.\label{Dirac}\end{equation}
This is rather surprising since the confluent hypergeometric equation with coefficients, (\ref{a-bis}) and (\ref{c-bis}), is completely equivalent to the Klein-Gordon equation
\begin{equation}
\frac{d^2\psi}{dr^2}+\left(\frac{\gamma E/\surd\left(m^2-E^2\right)}{r}-\frac{\ell(\ell+1)-\gamma^2}{r^2}-\fourth\right)\psi=0.\label{KG}
\end{equation}
It is commonly believed that (\ref{KG}) describes a spinless particle in a coulomb field, and is, therefore, not capable of describing the hydrogen atom since electrons have spin $\half$.\cite{Bethe}

Now the indicial equation for (\ref{KG}) about the origin has exponents
\begin{equation}
n_{1,2}=-\half\pm\surd\left[(\ell+\half)^2-\gamma^2\right], \label{indicial-bis}
\end{equation}
 which reduce to $n_1=\ell$ and $n_2=-(\ell+1)$ in the nonrelativistic limit [cf. Eq.~(\ref{indices})]. This would lead us to consider a solution to the indicial equation with only the former exponent. However, with $\ell=0$, the square root can become complex for $\gamma>\half$, and the solution would  diverge. A further complication  is that the exponents, (\ref{indicial-bis}), become complex for $\gamma>\half$ and $\ell=0$. With a complex exponents, the solutions near the origin would oscillate. Bethe\cite{Bethe} contends that the value of $Z$ that would be required to make the exponents complex would correspond to atoms whose radii are several times the Compton wavelength so as to invalidate the solution
 \begin{equation}
 \frac{e^{\pm i\gamma\ln r}}{\surd{r}},\label{soln}
 \end{equation}
  for small $r$. However true this may be, it would still make the energy levels, (\ref{Dirac}), complex, again making it unacceptable. We now address this in some detail.

\begin{figure}[htbp]
	\centering
		\includegraphics[width=0.25\textwidth]{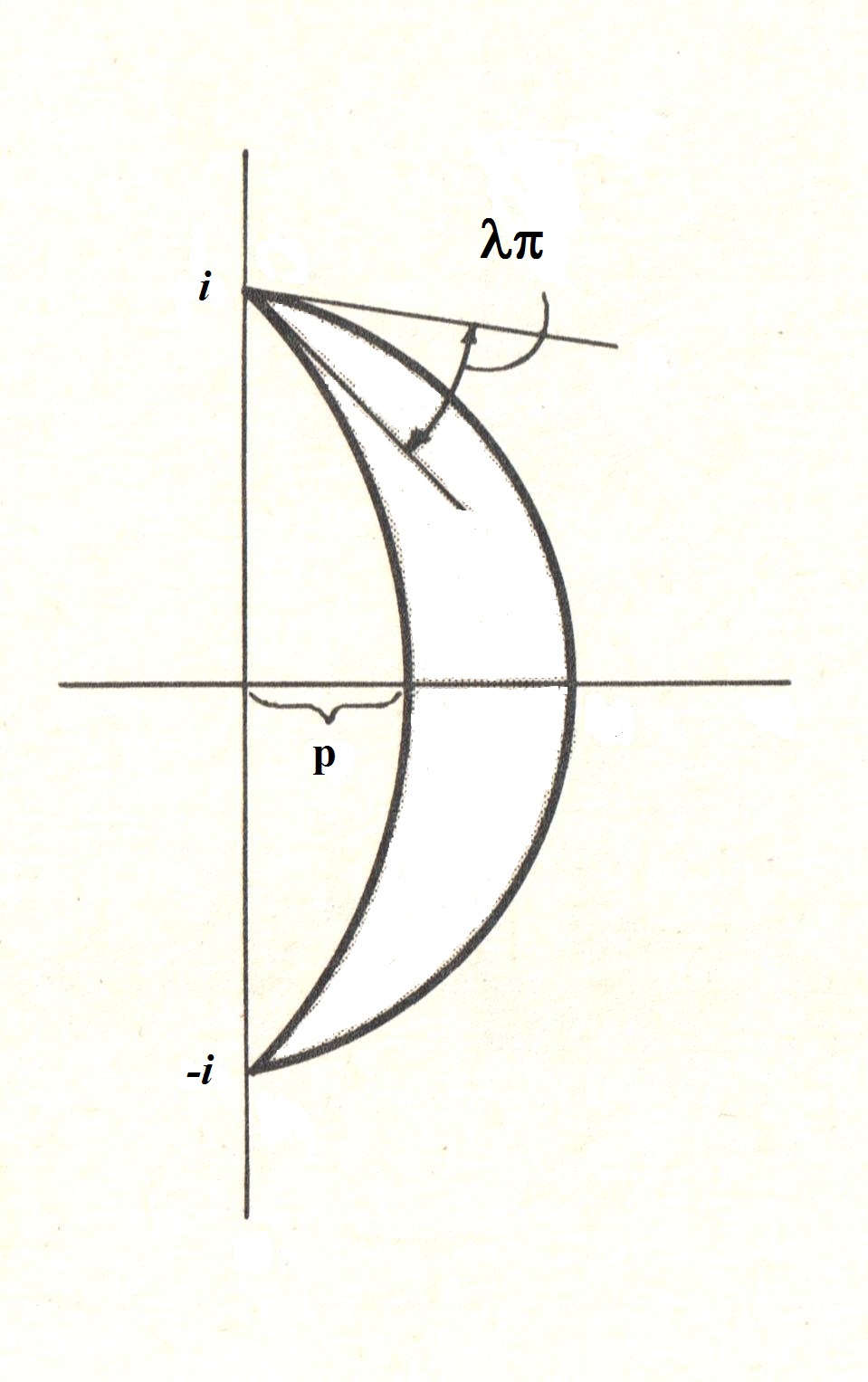}
	\caption{Crescent formed from intersecting circles making an angle $\lambda\pi$.}
	\label{crescent}
\end{figure}

  In the nonrelativistic limit, the ratio of the two solutions is $r^{\lambda}$, where $\lambda\pi=(2\ell+1)\pi$. The conformal mapping between the $z$ and $r$ planes is\cite{Forsyth*}
 \begin{equation}
 z=i\frac{r^{\lambda}+e^{-2i\cot^{-1}p}}{r^{\lambda}-e^{-2i\cot^{-1}p}}, \label{Forsyth}
 \end{equation}
 which upon solving for $r^{\lambda}$ becomes
 \begin{equation}
 e^{\lambda\ln r}=e^{-2i\cot^{-1} p}\left(\frac{z+i}{z-i}\right).\label{crescent-bis}
 \end{equation}
 The circles intersect at $-i$ and $i$, and $p$ is the distance from the smaller circle to the origin as shown in FIG.~\ref{crescent}. 
 These correspond to the branch points $r=0$ and $r=\infty$. Now, $\alpha=\cot^{-1}p=i\coth^{-1}(ip)$ and the right-hand side of (\ref{crescent-bis}) is  the cross ratio, 
 \begin{eqnarray} 
e^{\lambda\ln r}&=&e^{2(\coth^{-1}(ip)-\coth^{-1}(iz))}=\frac{p-i}{p+i}\cdot\frac{z+i}{z-i}\nonumber\\
&=&\frac{\left|\begin{array}{cc} p & 1\\ i & 1\end{array}\right|\cdot\left|\begin{array}{cc} z & 1 \\ -i & 1\end{array}\right|}{\left|\begin{array}{cc} p & 1\\ -i & 1\end{array}\right|\cdot\left|\begin{array}{cc} z & 1 \\ i & 1 \end{array}\right|}= \frac{(\cot\alpha-i)}{(\cot\alpha+i)}\cdot\frac{(\cot\beta+i)}{(\cot\beta-i)}\nonumber\\
&=& e^{-2i(\alpha-\beta)},\label{X-el}
 \end{eqnarray}
 where $p=\cot\alpha$ and $z=\cot\beta$. 
 Taking the logarithm of both sides of (\ref{X-el}) gives
 \begin{equation}
 -2i(\alpha-\beta)=\lambda\ln r, \label{X-el-bis}
 \end{equation}
 where the cross ratio is the distance between the points $(\cos\alpha,\sin\alpha,0)$ and $(\cos\beta,\sin\beta,0)$ with respect to the circular points at infinity, $(1,i,0)$ and $(1,-i,0)$ in the complex projective plane. They are called circular points at infinity because they lie on the complexification of every real circle. Both points satisfy the homogeneous equation,
 \begin{equation}
 Ax_1^2+Bx_2^2+2Cx_1x_3+2Dx_2x_3+Ex_3^2=0.\label{conic}
 \end{equation}
By specifying the line at infinity, $x_3=0$, the circular points then satisfy,
 \begin{equation}
 x_1^2+x_2^2=0,\label{invo-el}
 \end{equation}
 if $A=B=1$. The involution, (\ref{invo-el}), is \textit{elliptic} since it has imaginary circular points for its fixed elements. Equation (\ref{invo-el}) defines a pair of imaginary planes, $x_1\pm ix_2=0$.

 In the relativistic case, where $Z\alpha>>\ell+\half$  the indices, (\ref{indicial-bis}), become
 \begin{equation}
 n_{1,2}\simeq -\half\pm i\gamma. \label{indicial-tris}
 \end{equation}
 The ratio of the two solutions is $r^{2i\gamma}$, which is the crescent problem, but with an \textit{imaginary} angle. Rotate the crescent by $\pi/2$, and the circles \lq intersect\rq\ at $-1$ and $1$, which lie on the disc. Again specifying the line at infinity, $x_3=0$, the conic, (\ref{conic}), reduces to
\begin{equation}
x_1^2-x_2^2=0, \label{invo-h}
\end{equation}
for $A=-B=1$. Eq.~(\ref{invo-h}) is  a \textit{hyperbolic} involution, since it has $\pm1$ as its fixed elements, and defines a pair of real planes, $x_1\pm x_2=0$. 

With the crescent rotated, $p\rightarrow i\tilde{p}$, where $\tilde{p}\in\Re$, the cross ratio of the $4$ points, $(\tilde{p},1,0)$, $(z,1,0)$, $(1,1,0)$, and $(-1,1,0)$ is real and is given by
\begin{eqnarray} e^{2i\gamma\ln r}&=&e^{-2(\coth^{-1}\tilde{p}-\coth^{-1}z)}=\frac{\left|\begin{array}{cc} \tilde{p} & 1\\ 1 & 1\end{array}\right|\cdot\left|\begin{array}{cc} z & 1 \\ -1 & 1\end{array}\right|}{\left|\begin{array}{cc} \tilde{p} & 1\\ -1 & 1\end{array}\right|\cdot\left|\begin{array}{cc} z & 1 \\ 1 & 1 \end{array}\right|}\nonumber\\
&=&\frac{\tilde{p}-1}{\tilde{p}+1}\cdot\frac{z+1}{z-1}= \frac{(\coth\alpha-1)}{(\cot\alpha+1)}\cdot\frac{(\coth\beta+1)}{(\coth\beta-1)}\nonumber\\
&=& e^{-2(\alpha-\beta)},\label{X-hy}
 \end{eqnarray}
 where $\tilde{p}=\coth\alpha$ and $z=\coth\beta$. Hence,
 \begin{equation}
 i(\alpha-\beta)=\gamma\ln r. \label{X-hy-bis}
 \end{equation}
 The substitution is \textit{hyperbolic} since it has real fixed points which lie on the disc in terms of the homogeneous coordinates $(\cosh\alpha,\sinh\alpha,0)$ and $(\cosh\beta,\sinh\beta,0)$.
 
 A comparison (\ref{X-el-bis}) and (\ref{X-hy-bis}) shows that whereas in the former the angle is real while the logarithm of the cross ratio is pure imaginary,  and a many-valued quantity,  in the latter, the angle is imaginary while the logarithm  of the cross ratio is real. Thus, in both cases 
 \begin{equation}
 r=e^{i\vartheta},\label{ln}
 \end{equation} where the condition, $-\pi<\vartheta\le\pi$ determines the single-valued principal value of $\ln r$.  Both (\ref{X-el}) and (\ref{X-hy}) express angles in the form of a projective invariant. Taking the logarithm of both sides of the first equality in (\ref{X-el}) yields
 \begin{equation}
 \half\lambda\ln r=i\left(\cot^{-1}z-\cot^{-1}p\right). \label{cot}
 \end{equation}
 Introducing (\ref{ln}), and taking the cotangent of both sides yield
 \begin{equation}
\cot\left(\half\lambda\vartheta\right)=\frac{zp+1}{p-z}=\frac{\cot\alpha\cdot\cot\beta+1}{\cot\alpha-\cot\beta}=\cot(\beta-\alpha), \label{Mob-el}
 \end{equation}
 which, upon equating arguments, becomes
 \begin{equation}
 \vartheta=\frac{2}{\lambda}(\beta-\alpha). \label{vartheta}
 \end{equation}
 
 Likewise, taking the logarithm of the first equality in (\ref{X-hy}) gives
 \begin{equation}
 i\gamma\ln r=\coth^{-1}z-\coth^{-1}\tilde{p}. \label{coth}
 \end{equation}
Introducing (\ref{ln}) and taking the hyperbolic cotangent of both sides result in
\begin{equation}
\coth\left(\gamma\vartheta\right)=\frac{z\tilde{p}-1}{z-\tilde{p}}=\frac{\coth\alpha\cdot\coth\beta-1}{\coth\beta-\coth\alpha}=\coth(\alpha-\beta), \label{Mob-hy}
\end{equation}
or upon equating the arguments give
\begin{equation}
\vartheta=(\alpha-\beta)/\gamma.\label{vartheta-bis} 
\end{equation}

Whereas the M\"obius transformation in (\ref{Mob-el}),
\begin{equation}
z^{\prime}=\frac{z-p}{1+pz}, \label{Mob-el-bis}
\end{equation}
has imaginary fixed points, and is related to an imaginary circle, the M\"obius transform in (\ref{Mob-hy}),
 \begin{equation}
 z^{\prime}=\frac{z-\tilde{p}}{1-\tilde{p}z}, \label{Mob-hy-bis}
 \end{equation}
 has real fixed points and is the most general analytic function that maps the unit circle onto itself.\cite{Nehari} That is, for $z=e^{i\theta}$, $|z^{\prime}|=1$ since
\[|z-\tilde{p}|=|e^{i\theta}-\tilde{p}|=|e^{-i\theta}-\tilde{p}|=|e^{-i\theta}(1-\tilde{p}e^{i\theta})|=|1-\tilde{p}z|.\]

 \textit{Consequently, the small $r$ solution to the Klein-Gordon equation, (\ref{soln}), does not allow $r$ to be interpreted as a real, radial coordinate.} This is in contrast to the usual interpretation whereby the factor $[(\ell+\half)^2-\gamma]^{\half}$ introduces a fixed branch cut along the real axis from $\ell=-\half+\gamma$ to $\ell=-\half+\gamma$. When $\gamma>\half$ this cut \lq\lq overtakes the physical state $\ell=0$, and there is no consistent solution\rq\rq.\cite{Frautschi}

In summary,  the projective invariant in (\ref{X-el}) is pure imaginary, meaning that we are dealing with an elliptic substitution, while the angle is real, whereas the projective invariant in  (\ref{X-hy}) is real, meaning that hyperbolic substitutions that map the real circle into itself, while the angle pure imaginary.

Finally, we return to the small $r$ solution to the nonrelativistic Schr\"odinger equation and show that if $r$ is real, $\lambda$ must be complex. If we again rotate the crescent in FIG.~\ref{crescent} so that the vertices are at $-1$ and $+1$, while keeping $p$ real, we have
\begin{eqnarray}
r^{\lambda}&=&e^{-2i\cot^{-1}p}\left(\frac{z+1}{z-1}\right)\nonumber\\
&=&e^{2\left[\coth^{-1}z+\coth^{-1}(ip)\right]}
= e^{2(\beta-i\alpha)}. \label{mixed}
\end{eqnarray}
Eq.~(\ref{mixed}) shows that \textit{in order to interpret $r$ as a real, radial coordinate, the angle $\lambda$ must be complex, except in the limit as $p\rightarrow\infty$.} In that limit the crescent degenerates into a half circle, and (\ref{mixed}) becomes real. This is in contrast with conformal analysis which holds that $r$ is complex and the angle $\lambda$ is real and positive.

The corresponding M\"obius transform,
\begin{equation}
z^{\prime}=\frac{z+ip}{1+ipz}, \label{zprime-bis}
\end{equation}
has fixed points at $-1$ and $+1$, which are the vertices of the crescent. In the limit as $p\rightarrow\infty$, (\ref{zprime-bis}) becomes an inversion, $z^{\prime}=1/z$. This, again, maps circles into circles for which straight lines are regarded as circles that pass through the point of infinity. In other words, this transformation associates points in the interior of a unit circle with points exterior to it. So that it would appear that \textit{classical quantum mechanics emerges when conformality disappears, and the independent coordinate becomes real as well as the coefficient in the hypergeometric equation, (\ref{Kummer}).}

Eq.~(\ref{zprime-bis}) can be written as
\[
\frac{z^{\prime}+1}{z^{\prime}-1}=K\frac{z+1}{z-1}, \]
where the multiplier of the transform, 
\[
K=\frac{1+ip}{1-ip}=e^{i(\pi-2\alpha)},\]
 shows that the transformation is elliptic, i.e., $|K|=1$. The angle is constrained to the interval $0<\alpha<\pi/2$, and is determined by the height of the center of the smaller circle of the crescent from the origin of the $r$ plane. The transformation (\ref{mixed}) will therefore map the upper half of the $z$ plane onto the interior of the crescent in the $r$ plane with angles $\lambda\pi$.


\begin{thebibliography}{99}
\bibitem{Regge}T.~Regge, \textit{Nuovo Cimento} {\bf{14}} (1959) 951; \textit{ibid} {\bf{14}} (1960) 947; A.~Bottino, A.~M.~Longoni, and T.~Regge, \textit{ibid} {\bf{23}} (1962) 954.
\bibitem{Eden}R.~J.~Eden, P.~V.~Landshoff, D.~I.~Olive, and J.~C.~Polkinghorne, \textit{The Analytic S-Matrix} (Cambridge U. P., Cambridge, 1966).
\bibitem{BHL}B.~H.~Lavenda, \lq\lq The Khuri-Jones threshold factor as an automorphic function,\rq\rq\ arXiv:1206.3452.
\bibitem{Veneziano}G.~Veneziano, \textit{Nuovo Cimento} {\bf{A57}} (1968) 190.
\bibitem{Mott}N.~F.~Mott and H.~S.~W.~Massey, \textit{The Theory of Atomic Collisions\/}, 2nd ed. (Clarendon Press, Oxford, 1949), 52, eqn (26).
\bibitem{Gottfried}K.~Gottfried, \textit{Quantum Mechanics\/,} Vol. I: Fundamentals (Benjamin, New York, 1966), p. 148, eqn (4).
\bibitem{Singh}V.~Singh,\lq\lq Analyticity in the complex angular momentum plane of the Coulomb scattering amplitude,\rq\rq\ \emph{Phys. Rev.\/} {\bf{127}} (1962) 632--636.
\bibitem{LL}L.~D.~Landau and E.~M.~Lifshitz, \textit{Statistical Physics} 2nd ed. (Pergamon, Oxford, 1959), p. 152.
\bibitem{Forsyth}A.~R.~Forsyth, \textit{A Treatise on Differential Equations\/}, 6th ed. (Macmillan, London, 1956), p. 228.
\bibitem{Ovsienko}V.~Ovsienko and S.~Tabachnikov, \textit{Notices AMS} {\bf{56}} (2009) 34-36.
\bibitem{Choudhary}Compare with A.~R.~Choudhary, \lq\lq New relations between analyticity, Regge trajectories, Veneziano amplitude, and M\"obius transformations,\rq\rq\ arXiv: hep-th/0102019.
\bibitem{Forshaw}J.~R.~Forshaw and D. A. Ross, \textit{Quantum Chromodynamics and the Pomeron} (Cambridge U. P., Cambridge, 1997).
\bibitem{Lavenda}B.~H.~Lavenda, \lq\lq Errors in the bag model of strings, and Regge trajectories represent the conservation of angular momentum in hyperbolic space,\rq\rq\ arXiv:1112.4383.
\bibitem{Gray}J.~Gray, \textit{Linear Differential Equations and Group Theory from Riemann to Poincar\'e\/} (Birkh\"auser, Boston, 1986).
\bibitem{Ford}L.~R.~Ford, \textit{Automorphic Functions}, 2nd ed. (Chelsea Pub. Co., New York, 1929), p. 54.
\bibitem{Mumford}D.~Mumford, C.~Series, and D.~Wright, \textit{Indra's Pearls: The Vision of Felix Klein} (Cambridge U. P., Cambridge, 2002), p. 171.
\bibitem{Efimov}N.~V.~Efimov, \textit{Higher Geometry} (Mir, Moscow, 1980), p. 413.
\bibitem{Busemann}H.~Busemann and P.~J.~Kelly, \textit{Projective Geometry and Projective Metrics} (Academic Press, New York, 1953), p. 231.
\bibitem{Blatt}J.~M.~Blatt and V.~F.~Weisskopf, \textit{Theoretical Nuclear Physics} (Springer, New York, 1979), p. 330.
\bibitem{Omnes}R.~Omn\`es and M.~Froissart, \textit{Mandelstam Theory and Regge Poles} (Benjamin, New York, 1963), p.~27.
\bibitem{Bethe}H.~A.~Bethe, \textit{Intermediate Quantum Mechanics} (Benjamin, New York, 1964), p. 185.
\bibitem{Forsyth*}A.~R.~Forysth, \textit{Theory of Functions of a Complex Variable}, vol. II, 3rd ed. (Cambridge U. P., Cambridge, 1918), p. 685.
\bibitem{Nehari}Z.~Nehari, \textit{Conformal Mapping} (McGraw-Hill, New York, 1952), p. 164.
\bibitem{Frautschi}S.~C.~Frautschi, \textit{Regge Poles and S-Matrix Theory} (W. A. Benjamin, New York, 1963), p. 126.

\end{thebibliography}
\end{document}